\documentclass[11pt,a4paper]{article}

\usepackage[margin=1in]{geometry}
\usepackage{amsmath,amssymb}
\usepackage{graphicx}
\usepackage{booktabs}
\usepackage{array}
\usepackage{longtable}
\usepackage{float}
\usepackage{pdflscape}
\usepackage{needspace}
\usepackage{xcolor}
\usepackage{enumitem}
\usepackage[hidelinks]{hyperref}
\usepackage{microtype}

\newcolumntype{L}[1]{>{\raggedright\arraybackslash}p{#1}}

\title{AI-Research Agents in the Wild.\\
       From GitHub and arXiv to Regularities and Gaps.}
\author{%
  Aleksey Komissarov\thanks{Neapolis University, Pafos, Cyprus.
  Corresponding author; email to be finalised at submission.} \\
  \and
  Andrey Ustyuzhanin\thanks{Constructor Labs, Bremen, Campus Ring~1,
  28759, Germany; Constructor University, Bremen, Campus Ring~1,
  28759, Germany; Institute for Functional Intelligent Materials,
  National University of Singapore, 4 Science Drive~2, Singapore
  117544, Singapore. Email to be finalised at submission.}
}
\date{15 July 2026}

\begin{document}
\maketitle


\begin{abstract}
AI-research agents, also described as \emph{autoresearch} systems, combine
language models with tools, search, evaluation, and iterative modification of
research artifacts. Their public software ecology remains difficult to compare
because repositories, papers, benchmarks, libraries, and companion artifacts
are often counted as one population. We connect two agent-drafted and
human-adjudicated registries frozen on 10 June 2026: 139 canonical public
repository records and 101 papers about AI-assisted research systems, each
record carrying an evidence card anchored to its primary sources. Nine promoted design lineages
contain 59 canonical memberships among 52 repositories, with seven
repositories assigned to two lineages. We use this evidence to examine six
time-stamped candidate regularities about structural compatibility, iteration
cost, empty design cells, pattern strength, cross-lineage transfer, and pattern
growth. A prospective audit of 25 newly ingested repositories observed zero of
six specified trigger events. The outcome supplies bounded support for R1--R4.
Broader corpus evidence contradicted the original scope of R5, and a failed
directional prediction leaves R6 exploratory. The paper--repository graph
contains 212 distinct curator-assigned relatedness edges, each recording a
repository our curation coded as related to a paper: 64 of 101 papers carry at
least one such edge, those edges reach 23 of 139 repositories, and the six
most frequently linked repositories carry 120 of the 212 edges (56.6\%). A
full-text check of seven of those papers located a mention of the credited
repository for 3 of the 29 edges they carry, so the relation measures our
curation and not the papers' citation behavior. An identity audit
adjudicated against repository commit histories and public contributor records
confirms at least 18 paper authors who also author a canonical registry
repository, across 20 of 63 candidate author slots in 13 of 101 papers. A
byline-matched tier reaches 58 identities and 63 slots in 15 of 101 papers; it
is reported separately and more weakly, because 40 of those 63 slots come from
registry records that transcribe the paper's own author list, and the commit
histories of the repositories concerned cannot support at least 24 of the 63. These results
provide a source-grounded theory under prospective test, a curator-assigned
map of paper--repository relatedness, and explicit limits on claims about
public visibility and community structure.
\end{abstract}


\section{Introduction}
\label{hyb:intro}

Language-model systems now participate in research workflows by retrieving
literature, proposing hypotheses, generating or modifying code, running
experiments, evaluating outputs, and assembling reports. MLAgentBench
\cite{mlagentbench}, The AI Scientist \cite{ai-scientist}, ResearchAgent
\cite{research-agent}, and Agent Laboratory \cite{agent-laboratory} illustrate
different units of work and evaluation regimes. Their mechanisms combine
reasoning--action loops \cite{react}, self-critique \cite{reflexion}, search
over reasoning paths \cite{tree-of-thoughts}, and declarative optimization
\cite{dspy,mipro} with research-specific artifacts and acceptance criteria.

We use \emph{autoresearch} as an umbrella term for public software and
associated artifacts that implement, support, or evaluate
language-model-mediated research workflows. The public record includes
runnable agents, benchmarks, evaluation harnesses, libraries, coordination
substrates, curated lists, paper companions, and boundary artifacts. This
heterogeneity creates three measurement problems. First, repository counts
provide a public-artifact denominator; a runnable-agent denominator requires
role coding. Second, recurring
design decisions require a coding vocabulary that can be tested on later
cases. Third, a paper coded as related to a repository, a paper with
companion code, and a paper author who maintains a repository represent
different relations.

The study addresses three research questions:

\begin{description}[leftmargin=0.10\linewidth,labelwidth=0.08\linewidth,
  style=multiline,itemsep=3pt]
  \item[RQ1.] What design regularities recur in the public repository ecology,
  and how do nine promoted lineages relate to the 139-record registry?
  \item[RQ2.] What happens when six candidate regularities, with trigger
  conditions fixed before inspection, are applied to 25 newly ingested
  repositories?
  \item[RQ3.] How do curator-assigned relatedness, companion code, and
  confirmed authorship relations connect 101 papers to the repository
  registry?
\end{description}

The theory under test is concrete. R1 links structural compatibility to the
failure modes available to a system. R2 links rapid iteration to bounded,
programmatically scored artifacts. R3 concerns a persistent empty design cell
combining a vector signal with a language-model judge inside an optimization
loop. R4 asks why failure patterns lead the internal evidence ranking. R5
links unit of work to architecture lineage and tests how design patterns cross
lineages. R6 asks whether small pattern clusters grow quickly or remain small.
The time-stamped audit differentiates their current status: R1--R4 receive
bounded support, the broad form of R5 is contradicted, and R6 remains
exploratory after its directional companion prediction failed.

The paper makes three contributions:

\begin{enumerate}[label=(C\arabic*),leftmargin=*,itemsep=2pt]
  \item a design theory stated as six candidate regularities with explicit
  trigger observations and a prospective case-level audit;
  \item an evidence-grounded map that separates repository-registry size,
  lineage membership, artifact role, and recurring antipatterns; and
  \item distinct measurements of curator-assigned relatedness, companion code,
  and authorship overlap, including a corrected lower bound on participation
  across software and papers.
\end{enumerate}

For system builders, the regularities identify design constraints and
evaluation risks. For reviewers, the cross-corpus relations provide explicit
denominators for code-availability and community claims. For empirical
research on AI-assisted science, the prospective protocol records where the
theory survived, where its scope changed, and which questions remain open.

\begin{landscape}
\begin{figure}[H]
  \centering
  \includegraphics[width=0.98\linewidth]{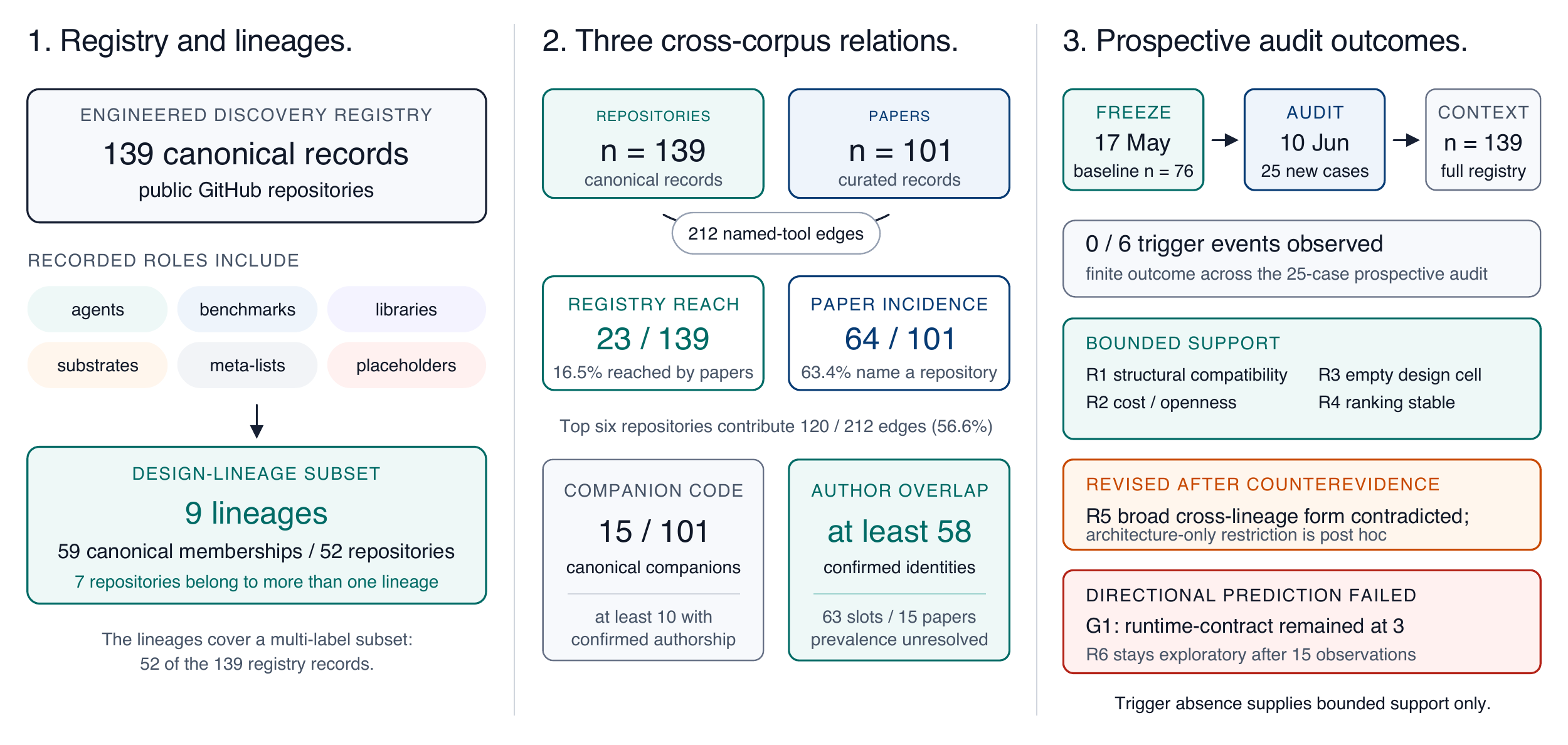}
  \caption{Study overview. The repository and paper registries support three
  connected analyses: lineage coverage, a 25-case prospective audit of six
  candidate regularities, and three paper--repository relations. The final
  panel distinguishes bounded support, a post-hoc restriction, and a failed
  directional prediction. Counts refer to the 10 June 2026 analytic freeze.}
  \label{hyb:fig:overview}
\end{figure}
\end{landscape}

\section{Corpus and Method}
\label{hyb:method}

The two registries were produced by the same procedure, which is
agent-drafted, human-adjudicated, and source-anchored. Each repository record
and each paper record received an evidence card written by a language-model
coding agent that was invoked once per record, worked from the primary
artifact, and filled a fixed card schema whose categorical fields draw on a
controlled vocabulary. The human authors set the admission scope, adjudicated
contested and unresolved cases, and reviewed the registries; they did not
hand-write the cards. The aggregate counts reported below are regenerated from
the structured fields of these cards, not from their prose. This is the
procedure disclosed in the ``Use of Generative AI'' section.

\subsection{Repository registry}

The repository registry contains 139 canonical public GitHub records after
three aliases are folded into their targets. Every admitted record carries a
recorded discovery channel, and the channels are unequal in yield. Of the 139
canonical records, 83 came from two existing curated lists (75 from the larger
seed list and 8 from a second list), 30 from synthesis passes that combined
snowball search over already inspected repositories with targeted GitHub
search, 15 from references in papers, 9 from a public benchmark leaderboard, 1
from a pull request opened against a curated list, and 1 from a direct
recommendation. No record entered through undirected GitHub search alone;
GitHub search is recorded as a technique used inside the synthesis passes and
not as a channel in its own right. Every admitted record received a
structured evidence card based on repository files, documentation, and linked
papers. The analytic freeze is 10 June 2026; later changes to live repositories
fall outside the reported counts.

The registry is an engineered discovery sample. Its records span research
agents, benchmarks, libraries, substrates, meta-resources, paper companions,
and artifacts that mark the category boundary. The count 139 therefore
describes a public artifact registry. Estimating the prevalence of runnable
agents requires a separate role-coding pass and an explicit sampling frame.

Repository inspection records canonical identity, discovery source, unit of
work, selection signal, judge type, loop structure, lineage evidence,
recurring patterns, and recurring failure modes. Source-derived observations
retain revision-pinned evidence in the public compendium. Aggregate counts are
regenerated from structured records.

The compendium holds 141 repository evidence cards against these 139 registry
records, and the two extra cards are enumerable. One documents a personal
development fork of MLAgentBench that was inspected to confirm redundancy and
then folded into its canonical target, so it keeps a card and no separate
registry row. The other documents AutoScientists, a repository inspected on 30
May 2026 whose registry row was never created; its evidence card is public, and
it is absent from every registry-derived count, including the 139-row
cost-versus-ambition matrix of Appendix~\ref{app:cost-ambition} and
Supplementary Table S1. The first difference is deliberate; the second is a
registry-maintenance error, stated here so that a reader comparing the two
lists finds exactly these two cards and no others. Neither affects a reported
number, because every count in this paper is computed over the 139 canonical
records.

\subsection{Paper registry and relation types}

The paper registry contains 101 curated records representing 98 unique
bibliographic artifacts. Three pairs share an arXiv identifier and retain
separate record IDs for provenance. The records contain bibliographic
metadata, authors, code-availability status, an optional companion-code URL,
and the canonical repositories our curation coded as related to it.
Twenty-five
records lacked accessible full text during curation, so paper-side link
coverage is an observed lower bound under the available-material protocol.

We distinguish three relations:

\begin{enumerate}[leftmargin=*,itemsep=2pt]
  \item A \emph{corpus-relatedness edge} connects a paper to a canonical
  repository that our curation coded as related to that paper. The field was
  introduced to cross-link the two registries, and the recorded basis for an
  edge ranges from an explicit citation in the paper to a curator's judgment
  that the paper and the repository occupy the same position in this corpus.
  These edges form a curator-assigned relatedness graph. We do not read them
  as a citation or naming graph, and the sample behind that decision is
  reported in the limitations.
  \item A \emph{companion-code edge} connects a paper to the canonical
  repository resolved from its own code URL. This relation measures associated
  public code.
  \item An \emph{authorship edge} connects a paper author to a canonical
  repository author. A \emph{byline-matched} edge requires only that the
  repository's registry record list the same name, on evidence from a companion
  repository, a project-author record, an adjudicated handle, or an equivalent
  identity source. A \emph{commit-adjudicated} edge additionally requires an
  author or committer identity in the repository's complete commit history.
  Only the commit-adjudicated class enters a lower bound: where a registry
  record's author field transcribes the byline of the paper being matched, the
  two sides of the comparison have one source and the match carries no
  repository-side information.
\end{enumerate}

Canonicalization resolves explicit aliases first and bare repository names
only when the basename identifies one canonical record. All current
relatedness labels resolve. Paper identifiers remain literal three-digit
strings, which preserves the 001--101 provenance range. One reproducibility
detail changes the headline counts and is therefore stated here: 100 of the
101 paper records store the relatedness field as an inline list and one stores
it as a block sequence, so a parser that accepts only the inline form returns
63 linked papers and 208 edges in place of 64 and 212. Our regeneration
accepts both forms.

\subsection{Lineages and coding primitives}

Nine promoted lineage records describe recurring relations among repositories.
Their membership lists declare 61 relations across 54 labels. Two labels are
external substrate anchors. Canonicalization leaves 59 memberships among 52
registry repositories, and seven repositories occur in two lineages. Lineage
membership is therefore a multi-label relation over a subset of the registry.
The recorded evidence combines code ancestry, shared infrastructure,
intellectual influence, and structural convergence.

Five coding primitives organize the design theory:

\begin{description}[leftmargin=0.23\linewidth,labelwidth=0.20\linewidth,
  style=multiline,itemsep=3pt]
  \item[A1. Unit of work.] The artifact produced or modified by one iteration.
  \item[A2. Selection signal.] The signal used to compare candidates. The
  coding vocabulary fixes five positions: a strict scalar, a band-tolerant
  scalar that admits a tolerance margin on secondary quantities, a vector
  compared on a frontier without scalarization, a relational score derived from
  pairwise comparison between candidates, and none for records that run no
  candidate comparison.
  \item[A3. Judge.] The acceptance mechanism. The coding vocabulary fixes four
  positions: programmatic, language-model, hybrid (a deterministic outer gate
  with an opt-in language-model inner check), and absent.
  \item[A4. Pattern.] A recurring design decision promoted to its own record
  under the compendium's recorded promotion rule. That rule fixes a minimum
  number of independently inspected artifacts, and every promoted record states
  the minimum it was promoted under.
  \item[A5. Antipattern.] A recurring implementation or evaluation failure with
  a named mechanism, promoted under the same rule as A4.
\end{description}

A2 and A3 are coded for every record, and their distributions are uneven
enough to state here. Over the 139 canonical records A2 is strict scalar in
85, none in 47, relational in 4, vector in 2, and band-tolerant in 1; A3 is
programmatic in 48, absent in 39, language-model in 33, and hybrid in 19. The
null position is the second most common value on both axes, and 38 of the 139
records take it on both at once. That is one concrete measure of the point
above: the registry counts public artifacts, and a substantial part of it runs
neither a candidate comparison nor an acceptance gate. The hybrid position
carries weight for R1 and R5 below, which both turn on whether a record's
acceptance gate is programmatic or language-model; 19 of 139 records are coded
as both, and a four-way scheme without a hybrid position would have forced
each of them silently onto one side. The vocabulary fixes no position for a
human acceptance mechanism, so human review is not separable in these counts.

The promotion rule behind A4 and A5 is not a single floor, and stating it
exactly changes how the pattern counts should be read. The compendium holds 54
promoted records: 22 architectures, 7 components, 16 coding patterns, and 9
antipatterns. Of those 54, 38 rest on two or more independently inspected
artifacts, and two of the 38 are recorded as promoted under a three-artifact
threshold. The remaining 16 rest on exactly one artifact and were promoted for
structural distinctness or for portability into a new system: 5 of the 22
architectures, 2 of the 7 components, 4 of the 16 coding patterns, and 5 of
the 9 antipatterns. Fifteen of those 16 declare the single-artifact status in
their own text, and one component record states the deviation outright,
recording that it was promoted on portability and utility merits and not at
the two-artifact threshold. Single-artifact promotion is a recorded policy of
the compendium and not a bookkeeping slip, and it has one direct consequence
for this paper: the pattern and antipattern totals reported below are counts
of promoted records and not counts of twice-confirmed regularities.

The two additions to the earlier seven antipattern records are \emph{mutable
evaluation code as a gaming surface} and \emph{calibration claimed but not
shipped}. The first is one of the 16 single-artifact records: it rests on a
single case study, a repository whose own documentation reports both the
gaming failure and the structural fix that ended it. Announcement-only paper
companions are coded as an artifact pattern, and tournament replacement of an
experiment remains a candidate observation.

\subsection{Time-stamped prospective protocol}

Six candidate regularities and their minimum trigger observations were
recorded on 17 May 2026 before the prospective batch was inspected. The
baseline registry contained 76 canonical records. Growth from 76 to 139
includes material accumulated outside the prospective batch; the explicit
case-level audit covers 25 repositories collected during June 2026.

The audit asks whether at least one qualifying trigger observation occurs in
those 25 cases. A trigger count of zero describes the finite audit batch.
Broader corpus evidence can still contradict a regularity, and a directional
prediction can fail when its paired trigger remains unobserved. We therefore
report trigger outcome, corpus-wide counterevidence, and post-hoc restrictions
as separate evidence layers.

\begin{table}[H]
  \centering
  \small
  \begin{tabular}{@{}L{0.06\linewidth}L{0.27\linewidth}L{0.46\linewidth}L{0.11\linewidth}@{}}
    \toprule
    \textbf{Code} & \textbf{Candidate regularity} & \textbf{Minimum trigger observation} & \textbf{Audit result} \\
    \midrule
    P1 & Structural compatibility constrains failure modes & One seed-line case with an uncalibrated language-model judge & 0 observed \\
    P2 & Iteration cost increases with artifact openness & One sub-10-minute iteration producing an open-ended artifact & 0 observed under measurements; 2 under configured bounds \\
    P3 & The vector-signal/language-model-judge optimization cell remains empty & One clean implementation of that cell: a per-candidate language-model judge, a retained vector with no reduction to a scalar before selection, and an iterated loop gated on the result & 0 observed \\
    P4 & Failure patterns lead the internal evidence ranking & An architecture pattern displaces them at the top of the ranking, scored under the imported four-lineage relation & 0 observed \\
    P5 & Unit of work constrains architecture lineage & One repository migrates between lineages across observation rounds & 0 observed \\
    P6 & Long-stagnant patterns rarely jump abruptly & One pattern stable at two instances for six rounds rises to at least five & 0 observed \\
    \bottomrule
  \end{tabular}
  \caption{Trigger conditions fixed before the 25-case audit. The result
  column reports observations in the audited batch. P4 is evaluated on the
  imported ranking, whose lineage term counts a four-lineage subset; under the
  nine promoted lineage records an architecture pattern holds the top position
  in two of the four weightings (Appendix~\ref{app:strength-sens}).}
  \label{hyb:tab:falsifiers}
\end{table}

\subsection{Artifact-level verification of the codings}
\label{hyb:method:verification}

Every count in this paper aggregates per-artifact codings recorded in the
evidence cards, so the codings themselves were checked against the artifacts
they describe. The check has two layers. The first resolves every
revision-pinned anchor: each \texttt{owner/repo@sha:path:line} reference
carried by the repository evidence cards was fetched at its pinned revision
and the cited line recorded verbatim. The second binds a load-bearing subset
of the codings to facts. A coding is load-bearing when a headline number, a
candidate regularity, or a printed table depends on it. Each such coding was
paired with a verbatim fact taken from the repository or the paper itself; the
evidence card was never accepted as evidence for its own coding. A negative
coding, for example \emph{uncalibrated} or \emph{no language-model judge},
required a stated exhaustive search: the revision searched, the paths and
patterns covered, and the result that nothing came back. A negative coding
with no stated search was recorded as not demonstrated. Every binding carries
one of five verdicts: demonstrated by a verbatim fact, demonstrated in a
weaker form, contradicted by the artifact, not demonstrated by the
discoverable evidence, or undecidable from the artifact.

The anchor layer resolves 3547 pinned references carried by the 141 repository
evidence cards, two of which lie outside the 139-record canonical registry. Of
those 3547 anchors, 3306 resolve to a verbatim source line at the pinned
revision, 132 name a file with no line number, 7 name a line past the end of
their file, and 102 name a path that does not resolve there. Twenty-seven of
those 102 are aimed at objects the anchor form cannot address at all: 15
commit-message fragments, 6 notebook cells, and 6 commit dates. One coverage
fact limits what the anchor layer establishes: 230 of the 3547 anchors are
cited inline beside the claim they support, and the remaining 3317 sit in a
page-level list, so the convention pins evidence to a card and not to a
sentence.

The binding layer covers 538 load-bearing codings over 54 artifacts, 44 of the
139 repository records and 10 of the 101 papers. Its outcome is asymmetric
across coding families. Contradicted bindings number 0 of 34 for judge
calibration, 0 of 44 for judge type, 0 of 44 for loop topology, 0 of 44 for
tool category, 0 of 44 for selection signal, 1 of 44 for mutability, 2 of 48
for lineage membership, 10 of 64 for pattern membership, and 10 of 59 for
iteration-cost coordinates. The four axes that supply every cell count,
lineage table, and candidate regularity in this paper, selection signal, judge
type, loop topology, and tool category, carry zero contradictions across their
176 bindings, and the empty-cell result of R3 is coded from two of them. The
calibration coding behind the leading failure pattern also survives: all 34
judge-calibration bindings resolve without contradiction, and 30 of the 34 are
demonstrated by a verbatim artifact fact or by a stated exhaustive search,
including the version split the finding depends on, in which The AI Scientist
\cite{ai-scientist} is coded validated on a shipped benchmark of 500
conference papers whose judge scores are compared with human ratings, and its
successor version is coded uncalibrated after that benchmark was dropped. The
two families that do not hold are the two that feed the weakest results
reported here: pattern membership supplies the supporting-repository counts of
the composite ranking (Appendix~\ref{app:strength-sens}), and the
iteration-cost coordinates supply the coverage matrix
(Appendix~\ref{app:cost-ambition}).

\subsection{Authorship audit and reproducibility}

The authorship analysis begins with 718 author slots across 101 papers and 661
distinct normalized display names. Display-name equality nominates a case for
review. Confirmation requires an explicit evidence source. Common names,
handle-only candidates, and affiliation conflicts remain unresolved.

Confirmed cases are then separated into two tiers, because the evidence behind
them is of two kinds. A byline-matched case pairs a paper author with a
canonical registry record whose own author field carries the same name. A
commit-adjudicated case additionally matches that person to an author or
committer identity in the repository's complete commit history, read across all
refs. Adjudication was run twice by different routes, once against public
contributor records and once against the histories themselves, and both results
are reported. Handle resemblance with no corroborating email or profile
evidence is recorded and counted separately.

We report lower bounds at three levels: confirmed identities, author slots,
and papers containing at least one confirmed repository author. A person on
several papers contributes one identity and several slots; a paper with
several confirmed repository authors contributes one paper. Lower-bound
language applies to the commit-adjudicated tier only. The byline-matched tier
is a count of attributed credit and bounds repository authorship in neither
direction, for the reason given with the result.

All graph counts use distinct paper--canonical-repository pairs. Percentages
retain their denominators. Supplementary Tables S1--S3 enumerate all 139
repositories, all 101 papers, and every observed named-tool, companion-code,
and confirmed-author relation, so each of those three relation counts can be
recounted from the article itself. The registries, evidence cards,
adjudication ledger, analysis programs, generated tables, figure source, and
manuscript source are deposited as the archival snapshot described under
``Data and Code Availability'' \cite{repo-metareview}.

\section{A Theory of Autoresearch Design}
\label{hyb:theory}

The theory describes empirical regularities in the observed public artifact
ecology. Each regularity has a mechanism, baseline evidence, a minimum trigger
observation, and a current status. The prospective protocol tests specified
events in 25 later cases. Corpus-wide descriptive evidence supplies context
and can expose counterexamples outside a narrow trigger.

\subsection{Six candidate regularities}

\paragraph{R1: structural compatibility.}
Structural commitments can remove specific failure modes from the available
design surface. The baseline records four zero co-occurrence pairs formed by
two structural conditions and two evaluation failures. P1 prospectively tests
one pair: seed-line systems combined with an uncalibrated language-model
judge. The only new seed-line case in the audit used a programmatic acceptance
mechanism, leaving that trigger unobserved. This is evidence for one observed
pair in one finite batch.

\paragraph{R2: iteration cost and artifact openness.}
Fast iteration is associated with bounded artifacts and programmatic scoring;
open-ended artifacts require more expensive judgment. P2 specifies a
sub-10-minute iteration that produces an open-ended artifact. Under manual
per-case coding, no audited case in the 25-repository batch combined the two
properties. Iteration cost and artifact openness lack independent double
coding, which limits the current inference.

Extending the coding to all 139 canonical repositories makes R2's status
conditional on a choice the earlier coding left implicit. Wall-clock evidence
divides into observed durations and configured bounds, and the two are not
interchangeable. Fourteen repositories publish a measurement; 56 publish only a
ceiling such as a timeout or a guard rail; 69 publish neither. If only
measurements count, the forbidden quadrant (iteration ${<}600$~s and artifact
open-ended) is empty, but the open-ended column then contains no measurement at
all across its 17 repositories, so the cell is empty for want of measurement
and not by measured absence. If configured bounds also count, as they must
if the bounded column is allowed to use them, two open-ended repositories bound
an iteration below ten minutes and P2 would have fired. We report both readings; adopting
the one that favours the hypothesis would be the easier move and the wrong one.
Appendix~\ref{app:cost-ambition} states the coding rule, both matrices, and the
unit recorded for every bound.

\paragraph{R3: persistent empty design cell.}
The coded corpus contains no clean optimization loop that retains a vector
signal and a language-model judge. At least 23 candidate systems landed in
adjacent designs. A common route reduces multiple judgments to a relational or
scalar tournament score. P3 observed no clean landing among the 25 audited
cases. The recorded trigger wording has three clauses, restated in
Table~\ref{hyb:tab:falsifiers}: a per-candidate language-model judge, a
retained vector with no reduction to a scalar before selection, and an
iterated loop gated on the result. A re-test run after the freeze found one
repository that satisfies the first two clauses and fails only the third, so
the third clause now carries the claim; Section~\ref{hyb:postfreeze} reports
that re-test, its protocol, and its other outcomes. The repeated adjacent
landings motivate a structural hypothesis that can be tested through
controlled scalar, vector, and tournament variants.

\paragraph{R4: pattern-strength asymmetry.}
Failure patterns lead the internal composite evidence ranking. The score
combines supporting repositories, papers, and lineages as
$s = 3\sqrt{n_{\text{tools}}} + n_{\text{papers}} + 2\,n_{\text{schools}}$,
with the top five patterns forming Tier S ($s \geq 12$). The rationale for the
three terms is stated here so that it does not depend on an external deposit:
the square root on $n_{\text{tools}}$ makes the step from one supporting
repository to four count for more than the step from nine to thirteen, because
the first question a replication count answers is whether the pattern recurs
at all; $n_{\text{papers}}$ enters linearly as corroboration from outside our
own reading of the code; and $n_{\text{schools}}$ carries weight two because
presence in more than one lineage is what separates a general pattern from an
idiomatic one. The weighting is one of several defensible choices. No architecture pattern displaced them
during the audit. Several supporting strength-increase thresholds were unmet,
so the stable rank is the bounded result. Appendix~\ref{app:strength-sens}
reports Tier S membership stability under three alternative weightings.
The ranking describes this corpus and coding system; independent data are
required to test generality.

\paragraph{R5: unit of work and cross-lineage transfer.}
The original theory proposed that design patterns would remain local to one or
two lineages. Broader corpus evidence contradicts that scope:
\emph{judge uncalibrated} occurs in four lineages. Architecture patterns
currently appear in at most two. Restricting R5 to architecture patterns
followed observation of the counterexample and is therefore post hoc. The
contrast between transferable evaluation failures and lineage-local
architectures remains a testable hypothesis.

\paragraph{R6: pattern-growth stability.}
The original heuristic proposed two growth modes: rapid accumulation of
instances or prolonged stagnation near the promotion threshold. P6 observed no
jump from a long-stagnant two-instance pattern to five instances. Its
directional companion prediction G1 expected the runtime-contract cluster to
gain a fourth instance; the count remained three across 15 recorded
observations. G1 failed, and R6 remains exploratory.

\begin{table}[H]
  \centering
  \small
  \begin{tabular}{@{}L{0.07\linewidth}L{0.27\linewidth}L{0.30\linewidth}L{0.27\linewidth}@{}}
    \toprule
    \textbf{Code} & \textbf{Baseline evidence} & \textbf{Prospective observation} & \textbf{Current status} \\
    \midrule
    R1 & Four recorded zero co-occurrence pairs & Sole new seed-line case lacked a language-model judge & Bounded support for one prospectively observed pair \\
    R2 & Fast cases use bounded artifacts; open-ended cases use costly evaluation & 0/25 combined both trigger properties; corpus-wide outcome depends on the evidence rule & Conditional: holds under measurements only, has two counterexamples under configured bounds \\
    R3 & At least 23 recorded adjacent or refused landings & 0/25 clean landings & Bounded support with tournament aggregation as an observed alternative; the closest post-freeze case fails only the loop clause \\
    R4 & A failure pattern leads the internal composite ranking under the imported four-lineage relation & No top-rank flip from a case observed in the audit & Conditional on the lineage relation: an architecture pattern leads under the nine promoted lineage records in two of the four weightings \\
    R5 & Unit of work proposed to constrain cross-lineage patterns & No migration trigger; one failure mode spans four lineages & Broad form contradicted; architecture-only form is post hoc \\
    R6 & Long-stagnant patterns proposed to remain small & No jump trigger; runtime-contract stayed at three & Exploratory; finite non-event and failed companion prediction \\
    \bottomrule
  \end{tabular}
  \caption{Current evidential status after separating the 25-case audit from
  broader corpus observations.}
  \label{hyb:tab:regularities}
\end{table}

\subsection{Prospective outcomes}

None of the six specified trigger events occurred among the 25 audited
repositories. The case-level interpretation is:

\begin{description}[leftmargin=0.08\linewidth,labelwidth=0.06\linewidth,
  style=multiline,itemsep=3pt]
  \item[P1.] One new seed-line case used programmatic acceptance; the
  uncalibrated language-model judge trigger was absent.
  \item[P2.] No audited case in the 25-repository batch combined sub-10-minute
  iteration with an open-ended artifact under manual coding. Coding all 139
  repositories afterwards makes the corpus-wide result rule-dependent: the
  forbidden quadrant is empty if only observed durations count, and holds two
  open-ended repositories if configured bounds count as well
  (Appendix~\ref{app:cost-ambition}).
  \item[P3.] No clean vector-signal/language-model-judge optimization loop was
  found; tournament aggregation occupied an adjacent design.
  \item[P4.] A failure pattern retained the leading position in the internal
  ranking under the imported four-lineage relation, which was computed at the
  76-record baseline and whose ranks two through five were not recomputed
  after the audit; several prespecified strength increases were unmet. Under
  the nine promoted lineage records an architecture pattern leads in two of
  the four weightings (Appendix~\ref{app:strength-sens}).
  \item[P5.] No lineage migration occurred in the audit. Independent corpus
  evidence contradicted the broad cross-lineage statement.
  \item[P6.] No long-stagnant pattern jumped to five instances. G1 failed after
  the runtime-contract count remained three across 15 observations.
\end{description}

The reproducible aggregate is zero observed trigger events among six tests in
25 cases. R1--R4 receive varying degrees of bounded support. R5 required a
post-hoc domain restriction. R6 remains exploratory. A statement of six
confirmations would exceed this evidence.

\subsection{What the theory currently explains}

Three mechanisms remain scientifically useful. Structural commitments appear
to constrain the failure surface. Rapid iteration currently co-occurs with
bounded, measurable artifacts. Evaluation failures can spread across lineages
that share no architecture. The persistent empty cell adds a fourth target:
tournament aggregation may provide the mechanism by which systems avoid
retaining vector judgments inside every optimization step, whether via
Elo-based co-evolution \cite{elo-evolve,deevo-tournament-prompts} or
social-choice ordinal aggregation \cite{soft-condorcet-optimization}. The
post-freeze re-test in Section~\ref{hyb:postfreeze} adds two further routes to
the same outcome, selection code shipped in a state no live path reaches and a
front constructed without an iteration that consumes it, so the mechanism is a
set of three routes and not one escape.

The theory also records its current boundary. The internal ranking measures
corpus support, with system performance outside its scope. R5's architecture
restriction is post hoc, and R6 has no successful directional prediction.
These limits turn the regularities
into an executable research program: future cases can populate a cell,
reverse a rank, cross a lineage boundary, or change a growth trajectory.

\subsection{Post-freeze re-test of the empty design cell}
\label{hyb:postfreeze}

R3 asserts a structural absence, and the registries that support it are frozen
on 10 June 2026. A structural claim invites a later check, so the empty cell
was re-tested on 26 July 2026 outside the frozen sample. The re-test collected
962 public repositories created after the freeze, retained 319 of them under a
mechanical filter, and passed those to ten language-model triage agents, each
required to quote a fact from the repository behind its verdict. The sweep is a
sample and not a census: the search queries return only their most recently
updated results. Sixty-nine of the 319 triaged candidates carried a trigger
flag. The highest-stakes of those flags were given to fifteen independent
re-checks, each instructed to reach the selection step by a different route and
to default to refutation, and 11 of the 15 refuted the claim they were given.
That ratio is the calibration figure to attach to any single-agent trigger
report, including the two that survived. This evidence is later than, and
separate from, the 25-case prospective audit, whose result is unchanged.

Four results follow. First, the closest recorded approach to the empty cell
satisfies two of the trigger's three clauses without qualification. In
\emph{jackysiupuichung/virtual-biotech-scientist}, six single-axis
language-model judges score each candidate against the incumbent, their
verdicts are retained as a six-element vector, and membership of the front is
gated on a conservative unanimity dominance predicate with no reduction to a
scalar before selection. The Copeland-style tally the repository also computes
runs strictly after selection and changes no membership. The third clause
fails: the front is constructed once, and its feedback signal is never
consumed. The trigger's phrase ``gates a loop'' therefore decides the case in a
way it did not at the freeze, because the case satisfies each of the trigger's
other clauses in the recorded wording. The next adjudication must state whether
a single-pass dominance selection over language-model-judged vectors fills the
cell, or whether the front must feed a later generation. The two readings
return opposite verdicts on the same repository.

Second, the emptiness has a third mechanism. The repository
\emph{smileformylove/XScientist} ships a complete and correct seven-dimension
Pareto pool that no live path reaches: its reviewer emits a flat score object
while the pool's only writer expects a nested one, so no candidate is ever
admitted. This was confirmed by executing the repository's own chain with its
feature flag forced on and observing zero admitted candidates and an empty
front. The cell therefore stays empty by three distinct routes: refused on
evaluation cost, escaped through tournament aggregation, and shipped
unreachable.

Third, the candidate recorded in advance as the one most likely to fill the
cell remains untestable. The AgenticOperator system described in one registry
paper was re-probed on 26 July 2026 across five repository-name variants and
the author's homepage; no code release was found.

Fourth, the same sweep produced one surviving case bearing on R1. The
repository \emph{arm00pv/nexus-swarm} accepts a mutated prompt on an
uncalibrated language-model judge's self-reported integer score, with no
programmatic co-gate, and an exhaustive search of its files finds no
calibration artifact. One question is open before the case can be counted
against R1's trigger: whether the mutated artifact lies on the causal path of
the scored scalar. If it does not, the case records a different failure, a gate
that fires while the mutation is inert, and not an instance of the R1 trigger.

The net effect on R3 is to raise its evidential standing. A structural claim
that was re-attacked after its freeze under a protocol built to refute it, and
that survived on one explicit clause, carries more information than the same
claim left frozen. What needs widening is the mechanism statement, from one
escape route to three. The emptiness itself stands, now with a named case that
shows what filling the cell would require.

\section{Public Artifact Ecology and Paper Links}
\label{hyb:links}

\subsection{Repository ecology and lineage coverage}

The nine promoted lineages cover 52 of 139 canonical repository records
(37.4\%). Their 59 memberships exceed the number of covered records because
seven repositories have two memberships. Eighty-seven registry records have
no promoted-lineage membership. The result is a structured, overlapping map
of a subset of the public registry.

The mixed artifact roles determine which corpus-wide claims are meaningful.
Repository ecology and the coded relatedness relation can use all 139
records. Runtime behavior, iteration cost, judge calibration, and agent
architecture require
role-specific denominators and case-level coding.

\subsection{Curator-assigned relatedness graph}

Sixty-four of 101 paper records carry at least one corpus-relatedness edge to
a canonical repository (63.4\%; nonparametric bootstrap 95\% CI
$[53.5, 72.3]$, $B=10{,}000$, seed 20260725). These edges reach 23 of 139
repository records (16.5\%; 95\% CI $[10.8, 23.0]$) and form 212 distinct
paper--repository pairs. Paper-side incidence and repository-side reach have
different denominators. Under adversarial imputation of the 25 paper records
whose full text was not retrievable, the paper-side rate is bounded in
$[63.4, 88.1]$; the observed value is a lower bound on the coded relation.

The coded relation is concentrated. Agent Laboratory \cite{agent-laboratory}
appears in 24 paper records, The AI Scientist \cite{ai-scientist} in 21,
MLAgentBench \cite{mlagentbench} in 20, AI-Researcher
\cite{hkuds-ai-researcher} in 19, NanoResearch \cite{nanoresearch} in 18,
and ResearchAgent \cite{research-agent} in 18. Together, these six
repositories contribute 120 of 212 edges (56.6\%). The identity of the
top-six set is stable under paper-level resampling: at least five of the
six members are retained in $91.5\%$ of $B=10{,}000$ replicates (mean
Jaccard $0.797$); the exact set recurs in $35.3\%$. The concentration is a
property of our coding. Its leading explanation is our own curation: which
repositories the registry admitted, and which repositories our curators knew
well enough to use as comparison anchors. Scientific influence, citation
inheritance, and annotation effort remain compatible secondary explanations;
the present coding does not separate them from curator familiarity.

\subsection{Companion code and authorship overlap}

Fifteen paper records have a companion-code URL that resolves to a canonical
repository. At least ten of those papers contain an author confirmed on the
companion repository.

The evidence ledger confirms at least 58 distinct paper authors who also
author one or more canonical registry repositories. They occupy 63 author
slots in 15 papers. Within the 64 papers that carry a relatedness edge, the
corresponding lower bounds are 57 identities, 62 slots, and 14 papers. In one
linked paper the coded relation points at a repository maintained by one of
its own authors, and six contain authors confirmed on other registry
repositories. Fifty linked
papers have no confirmed registry-author identity under the current ledger;
their cited-repository alignment remains unresolved.

\begin{table}[H]
  \centering
  \small
  \begin{tabular}{@{}L{0.49\linewidth}L{0.18\linewidth}L{0.25\linewidth}@{}}
    \toprule
    \textbf{Quantity} & \textbf{Value} & \textbf{Interpretation} \\
    \midrule
    Papers with at least one corpus-relatedness edge & 64/101 & observed paper-side incidence \\
    Repositories reached by relatedness edges & 23/139 & observed repository-side reach \\
    Distinct relatedness edges & 212 & curator-assigned; not citation-derived \\
    Edges incident on the six most linked repositories & 120/212 & 56.6\% concentration \\
    Papers with a canonical companion-code edge & 15/101 & associated code resolved \\
    Papers with a byline-matched author of that companion & at least 10/15 & companion authorship, byline-matched \\
    Paper authors adjudicated against a repository commit history & at least 18 & identity-level lower bound \\
    Author slots so adjudicated & at least 20/63 & slot-level lower bound \\
    Papers with a commit-adjudicated registry author & at least 13/101 & paper-level lower bound \\
    Byline-matched paper authors on a registry record & 58 & weaker tier; not a bound \\
    Byline-matched author slots & 63 & 40 of them from four byline-transcribing records \\
    Byline-matched slots excluded by commit histories & at least 24/63 & ceiling on the weaker tier \\
    Papers with a byline-matched registry author & 15/101 & weaker tier; not a bound \\
    \bottomrule
  \end{tabular}
  \caption{Three paper--repository relations at the analytic freeze.
  Authorship is reported in two tiers: commit-adjudicated rows are lower
  bounds; byline-matched rows count attributed credit and bound repository
  authorship in neither direction.}
  \label{hyb:tab:links}
\end{table}

\subsection{Interpretation}

The paper and repository registries overlap through curator-assigned
relatedness, companion artifacts, and people. Their strongest measured
structure is a concentrated relatedness graph: 64 of 101 curated papers are
coded as related to at least one canonical repository, and those edges reach
23 of 139 registry records. The graph is a real object and the concentration
is a real result; what it describes is our curation. Converting it into a
citation graph is a defined piece of work: a mention extraction over the 76
paper records with retrievable full text, admitting only edges with a
locatable mention in the paper, would yield a second graph that can be
compared against this one. The confirmed authorship
lower bound demonstrates direct participation across software and papers.
This revision reports lower bounds for that participation. Estimates of
overall prevalence and community structure require adjudicating the remaining
identity candidates and stratifying repositories by artifact role.

Private use lies outside both sampling frames. Its size and behavior are
unmeasured. Surveys, consent-based telemetry, or institutional sampling would
be required to study that population.

\section{Discussion}
\label{hyb:discussion}

\subsection{Scientific value and novelty}

The study advances empirical analysis of AI-assisted research systems in three
ways. It connects source-level repository evidence to a paper registry at the
level of canonical artifacts. It tests candidate design regularities against
cases whose triggers were fixed before inspection. It treats authorship as an
evidence-resolution problem with explicit lower bounds.

The coding scheme is reported together with a check of itself. That check is a
result and not only quality control: it establishes which aggregates in this
paper rest on codings a reader can verify at the source, and it names the two
coding families where that verification fails
(Section~\ref{hyb:method:verification}). An aggregate over a coding family
with no contradicted binding and an aggregate over a family with ten
contradictions in 64 bindings are different kinds of number, and this paper
prints which is which.

The original theoretical contribution survives in a more precise form. The
regularities specify mechanisms and future observations. The prospective
result is mixed: R1--R4 remain viable within a finite audit, R5 loses its broad
scope, and R6 loses its directional companion prediction. Reporting those
different outcomes preserves the theory's falsifiability.

The cross-corpus contribution also changes. The data support a concentrated
curator-assigned relatedness relation and confirmed author overlap. They leave
the magnitude and structure of community participation open, and they leave
the field's own citation behavior unmeasured. Separating relatedness,
companion-code, and authorship edges makes those boundaries explicit.

\subsection{Implications for system design}

R1 and R2 suggest that the acceptance mechanism and unit of work should be
designed together. Programmatic acceptance enables rapid iteration by
restricting the artifact to measurable outcomes. Open-ended artifacts require
additional judgment and calibration. Systems using language-model judges
should report calibration data, agreement, and sensitivity to prompt and
model choice.

R3 suggests a controlled experiment. Scalar, vector, relational, and
tournament variants can be compared on evaluation cost, ranking stability,
and search performance. Such a study would test whether tournament aggregation
is the structural escape route inferred from the observed adjacent designs.
Ordinal social-choice aggregation \cite{soft-condorcet-optimization}
provides a candidate mechanism for populating the empty cell without
retaining cardinal vector judgments; whether it does so at competitive
evaluation cost is an open empirical question.

R4 and R5 place evaluation infrastructure at the center of architecture
review. Judge-related failures occur across four lineages, whereas architecture
patterns currently remain within one or two. Shared evaluation components can
therefore transfer failure mechanisms across otherwise different systems.
Immutable evaluation code and shipped calibration artifacts are concrete
preventive measures.

R6 should be studied through a sequence of frozen snapshots. Survival or
event-history analysis could estimate time to promotion, stagnation, and
extinction without converting one failed prediction into confirmation.

\subsection{Implications for reviews of the field}

Repository and paper reviews should declare their unit of analysis. Counts of
repositories, runnable agents, papers, paper--repository edges, and people
answer different questions. Canonicalization and artifact-role coding should
precede prevalence estimates. Companion-code edges measure associated code. A
relatedness field assigned by curators measures the curators' map of the
field; a claim about intellectual visibility requires a mention extraction
from paper full text.

Authorship analyses require an explicit unresolved category, and they require
the evidence class to be named beside the count. A matcher built from a partial
collaborator directory measures that directory's coverage; the hit count
reported for one such matcher earlier in this project is not reproducible from
the current sources, it was never a measurement of overlap, and it is withdrawn
here. It is not comparable with the counts reported above, which are taken over
a different population under an explicit evidence policy, so the two must not
be read as before-and-after values of one quantity. What the present evidence
supports is a doubly adjudicated lower bound of 18 identities and a
byline-matched tier of 58 that the same commit histories cap from above. The
conclusion therefore depends strongly on the identity model and its evidence
policy, which is why both tiers are printed.

\subsection{Open problems}

The revised evidence changes the five main open problems:

\begin{description}[leftmargin=0.12\linewidth,labelwidth=0.10\linewidth,
  style=multiline,itemsep=3pt]
  \item[Q1.] The vector-signal/language-model-judge cell remains empty after at
  least 23 adjacent or refused landings. The post-freeze re-test in
  Section~\ref{hyb:postfreeze} narrows the question: one repository satisfies
  every clause of the recorded trigger except the iterated loop, so the answer
  now turns on whether a single-pass dominance selection over
  language-model-judged vectors counts as filling the cell. A controlled
  landing under an agreed reading of that clause would answer the
  structural-mechanism question.
  \item[Q2.] Symmetric multi-agent debate remains a separate empty-cell
  question and lacks a completed prospective test in this paper.
  \item[Q3.] Runtime-contract remained at three across 15 observations. The
  directional graduation prediction failed; the general growth heuristic
  remains open.
  \item[Q4.] Judge calibration failures recur across lineages. Cost, culture,
  deadlines, and tooling remain undistinguished as possible causes.
  \item[Q5.] The coded graph shows relatedness concentration and author
  overlap. A mention extraction over the 76 paper records with retrievable
  full text, complete entity resolution, and role-stratified denominators are
  needed for a community-structure estimate.
\end{description}

\subsection{Relation to prior work}

Surveys of agentic scientific discovery and deep-research agents organize
papers by planning, retrieval, tool use, synthesis, evaluation, research
stage, and human involvement
\cite{agentic-scientific-discovery-survey,deep-research-roadmap,
dr-survey-autonomous,automation-to-autonomy-survey,agentic-science-survey}.
Scientific-language-model surveys broaden the scope to
specialized models, multimodal scientific data, and process-oriented
evaluation \cite{sci-llm-survey}.
Point-system reports provide detailed accounts of individual architectures,
including MLAgentBench \cite{mlagentbench}, The AI Scientist
\cite{ai-scientist}, ResearchAgent \cite{research-agent}, and Agent Laboratory
\cite{agent-laboratory}; MLR-Copilot \cite{repo-G032} is a canonical exemplar
of the mlagentbench-derived lineage in our corpus.

The present study uses the public artifact and its recorded relations as the
unit of analysis. Repository inspection supplies implementation evidence;
canonicalization supplies stable identities; the paper registry supplies
curator-assigned relatedness, companion-code, and authorship relations; and
the prospective audit adds a temporal test of design claims. Together, these
elements complement capability-centered surveys with an evidence-grounded map.

\subsection{Limitations}

The repository registry is an engineered discovery sample shaped by search
terms, curated lists, snowballing, GitHub availability, and English-language
metadata. Global prevalence lies outside the design.

Artifact roles and several design variables lack independent double coding
across all records. The lineage relation combines ancestry, shared substrate,
intellectual influence, and structural similarity. A future ontology should
separate those relations and report inter-rater agreement.

The artifact-level check of the codings covered a load-bearing subset: 538
codings over 44 of the 139 repository records and 10 of the 101 papers. It is
not a re-coding of the corpus, and coding remained single-coder throughout, so
the check reports whether a coding is demonstrable from its source and not
whether a second coder would have assigned it. Two coding families produced
contradictions, and the results that depend on them are identified in
Section~\ref{hyb:method:verification}.

The paper registry is curated and includes 25 records without accessible full
text at the time of coding. Relatedness edges may be missed, annotation effort
may vary across papers, and an edge may also be recorded where the paper does
not mention the repository at all. The last direction was sampled directly.
Full text was retrieved for seven linked papers that carry 29 of the 212
edges, and each of the 29 credited repositories was searched for in the text
of the paper that credits it. Three of the 29 are locatable, one in prose and
two as reference-list entries; the remaining 26 have no locatable mention. On
that sample the relation is not a citation or naming relation, and the
concentration result is descriptive of our own coding.

The prospective audit contains 25 non-random cases. Its protocol was
time-stamped in project history; external preregistration governance was
absent. Several variables required judgment and lacked
independent double coding.

Authorship counts are lower bounds. Identity resolution remains incomplete for
common names, handle-only records, and ambiguous affiliations. The confirmed
cases establish overlap; prevalence and community structure remain open.

Repositories evolve after inspection. Revision-pinned evidence supports the
reproducibility of coded observations at the freeze. Later versions can change
architecture, documentation, and code availability. R3 asserts a structural
absence and was therefore re-tested outside the frozen sample;
Section~\ref{hyb:postfreeze} reports that re-test and its outcome.

\section{Conclusion}
\label{hyb:conclusion}

This study maps 139 canonical public repository records and 101 papers in the
ecology of AI-assisted research. Nine promoted lineages account for 59
memberships among 52 repositories and overlap on seven of them. The resulting
map separates the broad artifact registry from the smaller design-lineage
relation.

The six candidate regularities preserve the original paper's central
scientific program. Structural design choices constrain observed failure
modes; fast iteration co-occurs with bounded, programmatically scored
artifacts; the vector-signal/language-model-judge cell remains empty; and
failure patterns retain the leading internal evidence ranks. The broad
cross-lineage claim was contradicted, and the pattern-growth heuristic remains
exploratory after a failed directional prediction. These outcomes define the
next prospective tests.

The paper--repository graph contains 212 curator-assigned relatedness edges.
Sixty-four of 101 papers carry at least one, the graph reaches 23 of 139
repositories, and six repositories account for 56.6\% of its edges. A
full-text sample of 29 of those edges located a mention for three, so the
graph is a map of our curation and not a citation graph. The authorship audit
confirms
at least 58 paper authors who also author a registry repository. The public
record therefore combines a concentrated curator-assigned relatedness map
with direct participation across software and papers.

The scientific contribution lies in the conjunction of theory and evidence
discipline: regularities have explicit triggers, population claims have
declared denominators, and unresolved cases remain visible. This structure
supports future revisions without converting missing observations into
positive evidence.

\Needspace{10\baselineskip}
\section*{Use of Generative AI}
\addcontentsline{toc}{section}{Use of Generative AI}

Generative-AI coding agents, including Claude Code and OpenAI Codex, assisted
with candidate triage, source navigation, analysis-code development,
consistency checks, prose revision, and production of the vector overview
figure. They also drafted the corpus itself. Each of the 101 paper records was
produced by a scripted language-model pass over the paper, and each repository
record was drafted by a coding subagent reading the repository at a pinned
revision. Agent participation is recorded in the compendium's own commit
history: as of 26 July 2026, 63 of its 144 commits carry a co-authorship
trailer naming the assisting model, and two commits are authored by the bot
identity \texttt{copilot-swe-agent[bot]}.

The corpus was not read in a single pass. It grew through a sequence of
ingestion waves, each reading a small batch of repositories at a pinned
revision and writing per-tool records whose claims resolve to an
owner/repo@sha:path:line anchor. The registries reached 139 repository records
and 101 paper records and were frozen for analysis on 10 June 2026. Two later
exercises were built to stress the frozen findings rather than to enlarge the
corpus. The first was a prospective audit: the six specified triggers were
applied to 25 newly ingested repositories, and none of the six events was
observed. The second re-opened the one structural finding, the persistently
empty design cell reported as R3, after the freeze. It collected 962
repositories created since the freeze, retained 319 under a mechanical filter,
and passed those to ten language-model triage agents, each required to quote a
repository fact behind its verdict. Sixty-nine of the 319 carried a trigger
flag; the highest-stakes flags went to fifteen independent re-checks, each
instructed to reach the selection step by a different route and to default to
refutation, and eleven of the fifteen refuted the claim they were given. That
eleven-of-fifteen rate is the calibration a reader should attach to any
single-agent trigger report, including the two flags that survived it. The
re-test is reported in full in Section~\ref{hyb:postfreeze}. The sweep is a
sample and not a census, and its result is later than and separate from the
25-case audit, which it leaves unchanged.

Every headline count in this paper is recomputed from those agent-drafted
fields and from no independent second source: the 212 named-tool edges from
the \texttt{related\_corpus\_tools} field of the 101 paper records, the 718
author slots and 661 distinct author names from the \texttt{authors} field,
and the 59 lineage memberships from the \texttt{tools} field of the nine
lineage records. The pinned evidence anchors establish that a cited source
line says what it is quoted as saying; they do not make the coding built on
that line independent of the agent that wrote it, and no coding was
double-coded by a second annotator when it was written. The claim we defend is
therefore not that generated text and classifications were excluded as
evidence. It is that every coding was shipped as an auditable evidence card
carrying the anchor a reader can resolve, and that the codings have since been
re-derived against the artifacts they describe, with the outcome of that check
reported in this paper, including the coding families where the artifact
contradicts the coding. The human authors selected the research questions,
adjudicated scientific interpretations, reviewed the manuscript, and retain
responsibility for its content, including the parts an agent drafted first.

\section*{Data and Code Availability}
\addcontentsline{toc}{section}{Data and Code Availability}

The repository registry, paper registry, evidence cards,
identity-adjudication ledger, analysis programs, generated tables, figure
source, supplementary relation tables, and manuscript source are collected in
the companion compendium \cite{repo-metareview}. The identifier of record for
that material is a versioned archival snapshot with a persistent identifier,
deposited with the submitted version. The development repository from which
the snapshot is cut is a working mirror, and its public reachability on any
given date is not part of what this paper claims; the snapshot, and not a
repository URL, is the citation a reader should resolve. Supplementary
Tables S1--S3 reproduce the full 139-repository registry, the full 101-paper
registry, and every named-tool, companion-code, and confirmed-author relation,
so the population and relation counts reported in this paper can be recounted
from the article alone.

Three artifacts that a reader is most likely to want are stated in the article
itself, so that no reported result depends on reaching the deposit. The
composite pattern-strength weighting and the rationale for each of its three
terms are given with R4 in Section~\ref{hyb:theory}. The bootstrap procedure
is fully specified in Appendix~\ref{app:visibility-ci} by its replicate count,
its two resampling units, and its random seed, which is sufficient to
regenerate the intervals from the supplementary tables. The iteration-cost
coding rule, its coverage denominator, and the resulting matrix are given in
Appendix~\ref{app:cost-ambition}; the deposit adds the per-repository
evidence quotes.

\appendix
\section{Metric Definitions and Provenance}

\begin{table}[H]
  \centering
  \scriptsize
  \begin{tabular}{@{}L{0.25\linewidth}L{0.16\linewidth}L{0.22\linewidth}L{0.29\linewidth}@{}}
    \toprule
    \textbf{Result} & \textbf{Numerator} & \textbf{Denominator} & \textbf{Evidence object} \\
    \midrule
    Canonical repository registry & 139 records & 142 completed rows & canonical registry after three aliases are folded \\
    Promoted-lineage coverage & 52 repositories & 139 canonical repositories & membership relation from nine lineage records \\
    Promoted-lineage memberships & 59 memberships & 52 covered repositories & same relation; seven repositories have two memberships \\
    Prospective trigger outcomes & 0 events & 6 triggers applied to 25 cases & case-level prospective audit \\
    Post-freeze re-test of R3 & 0 clean landings, 1 two-clause case & 319 candidates retained from 962 post-freeze repositories & discovery sweep with ten triage agents and 15 adversarial re-checks \\
    Papers with a relatedness edge & 64 papers & 101 paper records & curator-assigned paper--repository edge table \\
    Repositories reached & 23 repositories & 139 canonical repositories & same edge table \\
    Relatedness concentration & 120 edges & 212 distinct edges & top-six repository edge counts \\
    Commit-adjudicated repository-author identities & at least 18 identities & 58 byline-matched identities & complete commit history of each implicated repository, two passes \\
    Commit-adjudicated repository-author slots & at least 20 slots & 63 byline-matched slots & same two passes \\
    Papers with a commit-adjudicated registry author & at least 13 papers & 101 paper records & same two passes \\
    Byline-matched repository-author identities & 58 identities & 661 normalized paper-author names & registry-record author fields; not a bound \\
    Byline-matched repository-author slots & 63 slots & 718 paper-author slots & registry-record author fields joined to paper authors \\
    Byline-matched slots from records transcribing a paper byline & 40 slots & 63 byline-matched slots & four registry records \\
    Byline-matched slots excluded by commit histories & at least 24 slots & 63 byline-matched slots & distinct non-bot committer identities per repository \\
    Papers with a byline-matched registry author & 15 papers & 101 paper records & registry-record author fields joined to papers \\
    Canonical companion-code edges & 15 papers & 101 paper records & resolved companion-code URLs \\
    Byline-matched companion authorship & at least 10 papers & 15 canonical companions & companion-record author fields \\
    \bottomrule
  \end{tabular}
  \caption{Denominators and evidence objects for headline results.}
\end{table}

\subsection{Pattern-strength weighting sensitivity}
\label{app:strength-sens}

The ranking summarised here carries two properties that bound its reading.
First, it was computed at the 76-record baseline registry and not at the 139
canonical repositories reported in the rest of the paper. A later spot check
against the enlarged corpus re-scored two of the five patterns, moving
\emph{judge-uncalibrated} from 16.5 to about 17.95 on at least 11 supporting
repositories instead of 8, and \emph{keep-or-revert} from 12.8 to about 13.22
on 14 instead of 13; the compendium records a full re-walk of the ranking
against the current corpus as deferred, so ranks two through five have not
been recomputed. Everything below is a sensitivity analysis on the 76-record
snapshot.

Second, the third input is not the lineage relation used elsewhere in this
paper. The imported score counts $c$ over a four-lineage subset (seed-line,
multi-agent-scientist, the Clune-lab open-endedness line, and the
benchmark-harness line), while every other lineage result reported here counts
membership over the nine promoted lineage records. The two relations give the
same $c$ for four of the five Tier S patterns. They differ for
\emph{keep-or-revert}: the imported score records one lineage at the 76-record
baseline, and the nine promoted records place its supporting repositories in
three, namely seed-line, the benchmark-harness line, and the
reinforcement-learning post-training line.

Table~\ref{tab:strength-sens} shows Tier S membership under four defensible
weightings of the same three inputs (tools $t$, papers $p$, lineages $c$).
Tier S is defined by the top-cluster gap in each variant. The baseline score
is $s_0 = 3\sqrt{t} + p + 2c$. Alternatives: linear tools
$s_1 = t + p + 2c$; unweighted lineages $s_2 = 2\sqrt{t} + p + c$;
paper-heavy $s_3 = 3\sqrt{t} + 2p + 2c$. The inputs are printed with the
scores so that every row can be recomputed.

\begin{table}[H]
  \centering
  \footnotesize
  \begin{tabular}{@{}L{0.30\linewidth}ccccccc@{}}
    \toprule
    \textbf{Pattern} & $t$ & $p$ & $c$ & $s_0$ & $s_1$ & $s_2$ & $s_3$ \\
    \midrule
    \textit{judge-uncalibrated}          &  8 & 4 & 2 & 16.5 & 16.0 & 11.7 & 20.5 \\
    \textit{staged-validation-before-judge} &  2 & 7 & 1 & 13.2 & 11.0 & 10.8 & 20.2 \\
    \textit{scoreboard}                  &  9 & 2 & 1 & 13.0 & 13.0 &  9.0 & 15.0 \\
    \textit{eval-discipline-regression}  &  4 & 3 & 2 & 13.0 & 11.0 &  9.0 & 16.0 \\
    \textit{keep-or-revert}              & 13 & 0 & 1 & 12.8 & 15.0 &  8.2 & 12.8 \\
    \midrule
    same pattern, $c$ recounted over the nine promoted lineages & 13 & 0 & 3 & 16.8 & 19.0 & 10.2 & 16.8 \\
    \bottomrule
  \end{tabular}
  \caption{Top-five pattern scores under four weightings, printed with the
  three inputs each score is computed from. The first five rows are the
  imported ranking at the 76-record baseline, with $c$ counted over the
  four-lineage subset. The final row recounts $c$ for the fifth pattern over
  the nine promoted lineage records and is not a sixth pattern. The identity
  of the top five is invariant across all four weightings and under both
  lineage relations; the within-cluster ordering and the leading position are
  not. Unweighted lineages rescales the whole tier and would require a
  proportionally lower threshold.}
  \label{tab:strength-sens}
\end{table}

Under the imported four-lineage relation an antipattern leads under all four
weightings. Under the nine promoted-lineage relation an architecture pattern
leads under the baseline weighting, 16.8 against 16.5, and under linear tools,
19.0 against 16.0, while the antipattern keeps the lead under unweighted
lineages, 11.7 against 10.2, and under the paper-heavy weighting, 20.5 against
16.8. Both relations are defensible; we report both and treat neither as a
settled ranking.

The movement comes from one supporting repository. Of the 13 repositories
credited to \emph{keep-or-revert}, nine are seed-line members, three belong to
no promoted lineage, and one, ML-Agent, supplies both the benchmark-harness
membership and the reinforcement-learning post-training membership. Counting
the same four-lineage subset against the current lineage records, in place of
the 76-record baseline, also raises $c$ from one to two through that one
repository and gives 14.8 under the baseline weighting, so part of the movement
is the snapshot and part is the choice of relation. The compendium page for
the pattern lists ML-Agent under refutations, with the note that it updates a
single policy in place by gradient descent and keeps no candidate pool to keep
or revert; the artifact-level re-check of pattern membership against primary
sources agrees, and it identifies NanoResearch as a repository that implements
the pattern and is absent from the supporting list. Adjudicating the contested
repository out and the missing one in leaves 13 supporting repositories
spanning two lineages under either relation, scoring 14.8 under the baseline
weighting and 17.0 under linear tools: the antipattern regains the lead under
the baseline weighting and the architecture pattern keeps it under linear
tools.

The top cluster contains two antipatterns, two architecture patterns, and one
component. That composition is preserved under all four weightings and under
both lineage relations. The leading position is not, and the composite is
presented as evidence-ranking within the corpus, not as an objective to be
optimised.

\subsection{Relatedness-graph uncertainty estimates}
\label{app:visibility-ci}

Confidence intervals on the paper- and repository-side relatedness rates use
nonparametric bootstrap with replacement, $B{=}10{,}000$ replicates and
random seed \texttt{20260725}. Paper-side resamples $101$ paper records;
repository-side resamples $139$ canonical repositories. Top-$6$ stability
counts, for each paper-level replicate, how many members of the observed
top-$6$ repository set (Agent Laboratory, The AI~Scientist, MLAgentBench,
AI-Researcher, NanoResearch, ResearchAgent) reappear when incidences are
recounted from the resample. The adversarial imputation bound treats all
$25$ paper records whose full text was not retrievable as carrying either
zero or at least one relatedness edge, yielding a floor equal to
the observed rate ($63.4\%$) and a ceiling of $88.1\%$. The parameters given above, $B{=}10{,}000$ replicates, the paper-side and
repository-side resampling units, and seed \texttt{20260725}, are the complete
specification of the procedure: the intervals can be regenerated from
Supplementary Tables S1--S3 with no code from the deposit. The driver that
produced the printed numbers and its outputs are in the archival
deposit~\cite{repo-metareview} under the data supplement described in the
``Data and Code Availability'' section.

\begin{table}[H]
  \centering
  \small
  \begin{tabular}{lrrl}
    \toprule
    Quantity & Point & 95\% CI & Notes \\
    \midrule
    Paper-side incidence & $63.4\%$ & $[53.5, 72.3]$ & resample 101 papers \\
    Repository-side reach & $16.5\%$ & $[10.8, 23.0]$ & resample 139 repositories \\
    Top-6 exact-set recurrence & $35.3\%$ & --- & mean Jaccard $0.797$ \\
    Top-6 ${\geq}5/6$ retained & $91.5\%$ & --- & paper-level resample \\
    Top-6 ${\geq}4/6$ retained & $99.8\%$ & --- & paper-level resample \\
    Paywalled imputation range & $63.4\%$ & $[63.4, 88.1]$ & 25 records \\
    \bottomrule
  \end{tabular}
  \caption{Bootstrap-derived uncertainty for the curator-assigned relatedness
  graph reported in the results section, $B{=}10{,}000$, seed $20260725$.}
  \label{tab:visibility-ci}
\end{table}

\subsection{Cost-versus-ambition coverage matrix}
\label{app:cost-ambition}

Regularity 2 asserts that no canonical repository combines sub-ten-minute
iteration with an open-ended artifact. This appendix reports the coverage of
that claim across the 139 canonical repositories. An \emph{iteration} is one
candidate-generation to acceptance-decision cycle within the tool's inner
loop.

The coding rule matters more than the coordinate, so it is stated first. Every
wall-clock coordinate falls into exactly one of three classes. A
\emph{measurement} is an observed duration: a results table, a run write-up, a
benchmark log, a README sentence reporting what a run took. A \emph{ceiling} is
a configured bound: a timeout default, a guard-rail rejection threshold, a kill
limit, a budget cap. The number describes a limit imposed, not a duration
observed. Everything else is \emph{unavailable}. Two auxiliary codes classify the
artifact as bounded, mixed, or open-ended and record the unit the source
actually uses, because a per-step, per-query or per-experiment figure bounds an
iteration only when that unit contains one. No numeric imputation is performed.

An earlier coding of this matrix did not separate the two classes, and applied
them asymmetrically: configured timeouts were accepted as wall-clock numbers in
the bounded column while the same class of evidence was recorded as unavailable
in the open-ended column. Re-coding all 139 rows against their pinned revisions
changes the picture in three ways. Fourteen rows carry a measurement, not the
six previously reported, and four of those six earlier anchors were ceilings or
values absent from the source they cited. Fifty-six rows carry a ceiling. The
open-ended column holds 17 repositories, not 7, so the previous coding
under-counted open-ended artifacts by more than half, which made the quadrant
look emptier than the evidence supports.

\begin{table}[H]
  \centering
  \small
  \begin{tabular}{lcccc}
    \toprule
    Iteration bucket & Bounded & Mixed & Open-ended & Unavailable \\
    \midrule
    \multicolumn{5}{l}{\emph{(a) Measurements only}} \\
    Under $600$~s               & 10 & 1  & 0  & 0  \\
    $600$~s to $1$~hour         & 0  & 1  & 0  & 0  \\
    $1$~hour to $1$~day         & 0  & 1  & 0  & 0  \\
    Over $1$~day                & 1  & 0  & 0  & 0  \\
    Iteration cost unavailable  & 61 & 24 & 17 & 23 \\
    \midrule
    \multicolumn{5}{l}{\emph{(b) Measurements and configured ceilings, unit respected}} \\
    Under $600$~s               & 23 & 4  & 2  & 0  \\
    $600$~s to $1$~hour         & 8  & 3  & 1  & 1  \\
    $1$~hour to $1$~day         & 11 & 6  & 1  & 1  \\
    Over $1$~day                & 2  & 0  & 1  & 0  \\
    Iteration cost unavailable  & 28 & 14 & 12 & 21 \\
    \bottomrule
  \end{tabular}
  \caption{Cost-versus-ambition coverage matrix over $139$ canonical
  repositories, under the two evidence rules. Rows partition by wall-clock
  iteration cost; columns partition by artifact openness. The
  under-$600$-second, open-ended cell is the forbidden quadrant predicted by
  Regularity 2: empty under (a), occupied by two repositories under (b). Each
  panel sums to $139$. The bottom bucket and the rightmost column are shown so
  that unavailable data stays visible.}
  \label{tab:cost-ambition}
\end{table}

Under rule (a) the forbidden quadrant is empty, but the open-ended column
contains no measurement at all: 0 of its 17 rows. The emptiness of that cell is
therefore an absence of measurement, not a measured absence, and the same
holds for the mixed column, which carries three measurements across 27 rows.
Rule (a) leaves the regularity consistent with every primary source and
supported by none of them in the region where it makes its prediction.

Under rule (b) the quadrant is occupied. Four open-ended repositories carry a
sub-$600$-second configured bound, and respecting the unit removes two of them:
one bound is a timeout on a single language-model API call and another on one
parallel branch, and neither constrains the duration of an iteration containing
it. Two survive. A deep-research agent sets a $500$-second default timeout on
one whole agent execution, so every iteration inside that execution completes
within it; a second sets a $270$-second query time limit over the whole run. Both
produce open-ended artifacts. Under this reading the prospective trigger P2 in
Table~\ref{hyb:tab:falsifiers} would have fired.

The honest statement of R2's status is therefore conditional, and we report it
that way; picking the rule that favours the hypothesis would be the easier move
and the wrong one. If only
observed durations count as evidence, the quadrant is empty and uninformative.
If configured bounds count, as they must if the bounded column is allowed to use
them, then two open-ended repositories bound an iteration below ten minutes and
the regularity has a counterexample. Converting this from a rule-dependent to a
settled result requires measured per-iteration durations in the open-ended
column, which no repository currently publishes. The per-repository coding, its
evidence quotes, the unit recorded for each bound, and the regeneration driver
are in the archival deposit~\cite{repo-metareview} under the data supplement
described in the ``Data and Code Availability'' section.

\section{Authorship Evidence Policy}

The identity audit accepts four evidence classes: authorship of a canonical
companion repository, an explicit contributor or project-author record, an
adjudicated repository profile or contribution identity, and a documented
cross-paper mapping with compatible affiliation evidence. A fifth path is
excluded: a registry record whose author field transcribes the byline of the
paper being matched. Where that field is an order-preserving prefix of the
paper's own author list, the match is uninformative, because both sides of the
comparison have one source. Four records of the current registry are of this
kind, and they carry 40 of the 63 byline-matched slots. Cases resting on that
path are counted in the byline-matched tier and are excluded from every lower
bound.

Each accepted case is adjudicated a second time against the complete commit
history of the repository, read across all refs and including history not
reachable from the default branch. A case is commit-adjudicated when the person
matches an author or committer identity in that history. Handle resemblance
with no corroborating email or profile evidence is reported separately and does
not enter a bound. The same histories supply an upper check on the
byline-matched tier: a repository with fewer distinct non-bot committer
identities than the slots charged to it cannot support the surplus.

Exact display-name equality nominates a case for review; confirmation requires
an explicit source. Conflicting affiliations, common names, and unresolved
handles remain open. The commit-adjudicated lower bound can increase as
evidence is added. The byline-matched count bounds repository authorship in
neither direction. Unresolved cases remain unclassified with respect to
community alignment.


\clearpage
\newgeometry{margin=0.55in}
\begin{landscape}

\begingroup
\setcounter{table}{0}
\renewcommand{\thetable}{S\arabic{table}}
\setcounter{section}{0}
\renewcommand{\thesection}{S\arabic{section}}
\setlength{\LTpre}{4pt}
\setlength{\LTpost}{8pt}
\setlength{\parindent}{0pt}
\setlength{\parskip}{4pt}

\begin{center}
  {\LARGE\bfseries Supplementary Material\par}
  \vspace{8pt}
  {\Large\bfseries AI-Research Agents in the Wild:\par}
  {\Large\bfseries From GitHub and arXiv to Regularities and Gaps\par}
  \vspace{12pt}
  {\large Aleksey Komissarov \qquad Andrey Ustyuzhanin\par}
  \vspace{8pt}
  {\normalsize Analytic freeze: 10 June 2026\par}
\end{center}

\section*{Scope and relation definitions}

This supplement enumerates every record used in the repository and paper
registries: 139 canonical public GitHub repositories and 101 paper records.
The paper records resolve to 98 unique bibliographic artifacts. Three pairs
share an arXiv identifier: P031/P033, P061/P068, and P081/P088. Both record
identifiers remain visible to preserve the curated provenance.
Repository identifiers \texttt{G001}--\texttt{G139} are assigned
lexicographically within this analytic freeze. Paper identifiers
\texttt{P001}--\texttt{P101} preserve the curated paper-record identifiers.

Three relations are reported separately. A \emph{named-tool} relation means
that a paper names, cites, or discusses a canonical repository. A
\emph{companion-code} relation means that the paper's own code URL resolves to
a canonical repository. A \emph{confirmed-author} relation requires explicit
evidence that at least one paper author also authors the repository. Blank
cells record an unobserved or unresolved relation under the stated protocol;
they provide no evidence of population separation.

The tables are generated from the frozen registries and adjudicated identity
ledger. Each repository and paper row carries a citation. The corpus-source
bibliography follows the tables, and machine-readable versions accompany the
article.

\section{Complete canonical repository registry}
\fontsize{9pt}{10.2pt}\selectfont
\setlength{\tabcolsep}{3pt}
\begin{longtable}{@{}L{0.48in}L{2.45in}L{1.45in}L{2.85in}L{1.15in}L{1.55in}@{}}
\caption{All 139 canonical public repository records at the analytic freeze. Paper identifiers in the last three columns refer to named-tool, companion-code, and confirmed-author relations, respectively.}\label{tab:supp-repositories}\\
\toprule
\textbf{ID} & \textbf{Canonical repository} & \textbf{Promoted lineage(s)} & \textbf{Named by papers} & \textbf{Companion of} & \textbf{Author overlap} \\
\midrule
\endfirsthead
\multicolumn{6}{l}{\footnotesize\emph{Table \thetable\ continued}}\\
\toprule
\textbf{ID} & \textbf{Canonical repository} & \textbf{Promoted lineage(s)} & \textbf{Named by papers} & \textbf{Companion of} & \textbf{Author overlap} \\
\midrule
\endhead
\midrule
\multicolumn{6}{r}{\footnotesize Continued on next page}\\
\endfoot
\bottomrule
\endlastfoot
\texttt{G001} & \href{https://github.com/ac-rad/organa}{\nolinkurl{ac-rad/organa}}\allowbreak{}\hspace{0.2em}\cite{repo-G001} & physical-robotic-automation & \textemdash & \textemdash & \textemdash \\
\texttt{G002} & \href{https://github.com/activeloopai/autoresearch_deeplake_swarm}{\nolinkurl{activeloopai/autoresearch_deeplake_swarm}}\allowbreak{}\hspace{0.2em}\cite{repo-G002} & \textemdash & \textemdash & \textemdash & \textemdash \\
\texttt{G003} & \href{https://github.com/affaan-m/everything-claude-code}{\nolinkurl{affaan-m/everything-claude-code}}\allowbreak{}\hspace{0.2em}\cite{repo-G003} & \textemdash & \textemdash & \textemdash & \textemdash \\
\texttt{G004} & \href{https://github.com/Agent-Analytics/autoresearch-growth}{\nolinkurl{Agent-Analytics/autoresearch-growth}}\allowbreak{}\hspace{0.2em}\cite{repo-G004} & \textemdash & \textemdash & \textemdash & \textemdash \\
\texttt{G005} & \href{https://github.com/agent-on-the-fly/InfoSeeker}{\nolinkurl{agent-on-the-fly/InfoSeeker}}\allowbreak{}\hspace{0.2em}\cite{repo-G005} & dr-deep-research-agent & \textemdash & \texttt{P062} & \texttt{P062} \\
\texttt{G006} & \href{https://github.com/Agent-RL/ReCall}{\nolinkurl{Agent-RL/ReCall}}\allowbreak{}\hspace{0.2em}\cite{repo-G006} & rl-post-training-as-autoresearch & \textemdash & \textemdash & \textemdash \\
\texttt{G007} & \href{https://github.com/agentmemoryworld/awesome-agent-memory}{\nolinkurl{agentmemoryworld/awesome-agent-memory}}\allowbreak{}\hspace{0.2em}\cite{repo-G007} & \textemdash & \textemdash & \texttt{P049} & \textemdash \\
\texttt{G008} & \href{https://github.com/ai-agents-2030/awesome-deep-research-agent}{\nolinkurl{ai-agents-2030/awesome-deep-research-agent}}\allowbreak{}\hspace{0.2em}\cite{repo-G008} & \textemdash & \textemdash & \textemdash & \texttt{P001},\allowbreak{} \texttt{P062} \\
\end{longtable}
\section{Complete paper registry}
\begin{longtable}{@{}L{0.48in}L{2.90in}L{0.90in}L{1.05in}L{2.55in}L{0.80in}L{1.15in}@{}}
\caption{All 101 paper records at the analytic freeze. Repository IDs resolve through Table~S1. The author column lists repositories with at least one evidence-confirmed paper author; bracketed values give the number of confirmed author slots for that paper--repository pair.}\label{tab:supp-papers}\\
\toprule
\textbf{ID} & \textbf{Paper title} & \textbf{Year / arXiv} & \textbf{Code / access status} & \textbf{Named repositories} & \textbf{Companion} & \textbf{Author links} \\
\midrule
\endfirsthead
\multicolumn{7}{l}{\footnotesize\emph{Table \thetable\ continued}}\\
\toprule
\textbf{ID} & \textbf{Paper title} & \textbf{Year / arXiv} & \textbf{Code / access status} & \textbf{Named repositories} & \textbf{Companion} & \textbf{Author links} \\
\midrule
\endhead
\midrule
\multicolumn{7}{r}{\footnotesize Continued on next page}\\
\endfoot
\bottomrule
\endlastfoot
\texttt{P001} & Memento: Fine-tuning LLM Agents without Fine-tuning LLMs\allowbreak{}\hspace{0.2em}\cite{paper-P001} & 2025 & code shipped & \texttt{G051},\allowbreak{} \texttt{G064},\allowbreak{} \texttt{G097},\allowbreak{} \texttt{G111} & \texttt{G083} & G083 [11]; G008 [1] \\
\texttt{P002} & DeepResearchArena\allowbreak{}\hspace{0.2em}\cite{paper-P002} & \textemdash & full text unavailable & \textemdash & \textemdash & \textemdash \\
\texttt{P003} & MoNaCo\allowbreak{}\hspace{0.2em}\cite{paper-P003} & \textemdash & full text unavailable & \textemdash & \textemdash & \textemdash \\
\texttt{P004} & DRA-MedicalAI\allowbreak{}\hspace{0.2em}\cite{paper-P004} & \textemdash & full text unavailable & \textemdash & \textemdash & \textemdash \\
\texttt{P005} & DeepResearch: A Survey of Autonomous Research Agents\allowbreak{}\hspace{0.2em}\cite{dr-survey-autonomous} & 2025 & paper only & \texttt{G064},\allowbreak{} \texttt{G097},\allowbreak{} \texttt{G110},\allowbreak{} \texttt{G111},\allowbreak{} \texttt{G129} & \textemdash & \textemdash \\
\texttt{P006} & A Survey of Scientific Large Language Models: From Data Foundations to Agent Frontiers\allowbreak{}\hspace{0.2em}\cite{sci-llm-survey} & 2025; 2508.21148 & code cited & \texttt{G111} & \texttt{G093} & G093 [10] \\
\texttt{P007} & FutureX: An Advanced Live Benchmark for LLM Agents in Future Prediction\allowbreak{}\hspace{0.2em}\cite{paper-P007} & 2025; 2508.11987 & demo only & \texttt{G097},\allowbreak{} \texttt{G111},\allowbreak{} \texttt{G032},\allowbreak{} \texttt{G116} & \textemdash & \textemdash \\
\texttt{P008} & A Survey on Agentic Multimodal Large Language Models\allowbreak{}\hspace{0.2em}\cite{paper-P008} & 2025; 2510.10991 & code cited & \texttt{G079},\allowbreak{} \texttt{G032},\allowbreak{} \texttt{G116} & \texttt{G050} & G050 [10] \\
\end{longtable}
\section{All observed paper--repository relations}
\begin{longtable}{@{}L{0.48in}L{0.48in}L{3.05in}L{2.55in}L{0.75in}L{2.65in}@{}}
\caption{Union of all observed paper--repository relations. N denotes a named-tool edge, C a canonical companion-code edge, and A an evidence-confirmed paper-author/repository-author relation. Multiple letters on one row describe distinct evidence for the same pair.}\label{tab:supp-relations}\\
\toprule
\textbf{Paper} & \textbf{Repo} & \textbf{Paper title} & \textbf{Canonical repository} & \textbf{Relation} & \textbf{Confirmed paper author(s)} \\
\midrule
\endfirsthead
\multicolumn{6}{l}{\footnotesize\emph{Table \thetable\ continued}}\\
\toprule
\textbf{Paper} & \textbf{Repo} & \textbf{Paper title} & \textbf{Canonical repository} & \textbf{Relation} & \textbf{Confirmed paper author(s)} \\
\midrule
\endhead
\midrule
\multicolumn{6}{r}{\footnotesize Continued on next page}\\
\endfoot
\bottomrule
\endlastfoot
\texttt{P001} & \texttt{G051} & Memento: Fine-tuning LLM Agents without Fine-tuning LLMs & \href{https://github.com/HKUDS/AI-Researcher}{\nolinkurl{HKUDS/AI-Researcher}} & N & \textemdash \\
\texttt{P001} & \texttt{G064} & Memento: Fine-tuning LLM Agents without Fine-tuning LLMs & \href{https://github.com/JinheonBaek/ResearchAgent}{\nolinkurl{JinheonBaek/ResearchAgent}} & N & \textemdash \\
\texttt{P001} & \texttt{G083} & Memento: Fine-tuning LLM Agents without Fine-tuning LLMs & \href{https://github.com/Memento-Teams/Memento}{\nolinkurl{Memento-Teams/Memento}} & CA & Guchun Zhang; Huichi Zhou; Jun Wang; Ka Yiu Lee; Kin Hei Lee; Kun Shao; Linyi Yang; Siyuan Guo; Xue Yan; Yihang Chen; Zihan Wang \\
\texttt{P001} & \texttt{G097} & Memento: Fine-tuning LLM Agents without Fine-tuning LLMs & \href{https://github.com/OpenRaiser/NanoResearch}{\nolinkurl{OpenRaiser/NanoResearch}} & N & \textemdash \\
\texttt{P001} & \texttt{G111} & Memento: Fine-tuning LLM Agents without Fine-tuning LLMs & \href{https://github.com/SamuelSchmidgall/AgentLaboratory}{\nolinkurl{SamuelSchmidgall/AgentLaboratory}} & N & \textemdash \\
\texttt{P001} & \texttt{G008} & Memento: Fine-tuning LLM Agents without Fine-tuning LLMs & \href{https://github.com/ai-agents-2030/awesome-deep-research-agent}{\nolinkurl{ai-agents-2030/awesome-deep-research-agent}} & A & Yihang Chen \\
\texttt{P005} & \texttt{G064} & DeepResearch: A Survey of Autonomous Research Agents & \href{https://github.com/JinheonBaek/ResearchAgent}{\nolinkurl{JinheonBaek/ResearchAgent}} & N & \textemdash \\
\texttt{P005} & \texttt{G097} & DeepResearch: A Survey of Autonomous Research Agents & \href{https://github.com/OpenRaiser/NanoResearch}{\nolinkurl{OpenRaiser/NanoResearch}} & N & \textemdash \\
\end{longtable}

\endgroup

\end{landscape}
\restoregeometry

\clearpage

\begingroup

\small

\sloppy

\renewcommand{\refname}{Corpus sources}

\endgroup

\end{document}